\documentclass[twocolumn,showpacs,superscriptaddress,prl]{revtex4-2}

\usepackage{bm}
\usepackage{graphicx}
\usepackage[normalem]{ulem}
\usepackage{ragged2e} 
\usepackage[utf8]{inputenc}
\usepackage{amsmath,empheq} 
\usepackage{xcolor}
\usepackage{physics}
\usepackage{gensymb}
\usepackage{float}
\usepackage{hyperref}
\usepackage{natbib}
\usepackage{newtxtext} 
\usepackage[varbb,varg]{newtxmath} % scaled=1.005
\begin{document}

\title{Orbital Competition in Bilayer Graphene's Fractional Quantum Hall Effect}
\author{Bishoy M. Kousa}
\email{bishoy.kousa@utexas.edu}
\affiliation{Department of Physics, University of Texas at Austin, Austin, TX 78712, USA}
\email{bishoy.kousa@utexas.edu}
\author{Nemin Wei}
\affiliation{Department of Physics, University of Texas at Austin, Austin, TX 78712, USA}
\affiliation{Department of Physics, Yale University, New Haven, CT 06520, USA}
\author{Allan H. MacDonald}
\email{macd@physics.utexas.edu}
\affiliation{Department of Physics, University of Texas at Austin, Austin, TX 78712, USA}

\date{\today}

\begin{abstract}
The lowest Landau level of bilayer graphene has an octet of internal degrees of freedom,
composed from spin, valley and orbital two-level systems.  Dominance of $n=0$ orbitals over 
$n=1$ orbitals in low energy quantum fluctuations leads to distinct fractional quantum Hall characteristics compared dominance of $n=1$ over $n=0$. 
The competition between $n=0$ and $n=1$ orbitals depends sensitively on
particle-hole asymmetry and on Lamb shifts due to exchange interactions with the negative energy sea, which
must be accounted for simultaneously in assessing the orbital competition.
We identify the circumstances under which $n=1$, 
which supports strong even-denominator FQH states with non-abelian quasiparticles, 
emerges robustly as the low-energy Landau level.  
\end{abstract}

\maketitle

\textit{Introduction} -- The lowest Landau level (LL) of bilayer graphene has 
an octet of discrete degrees of freedom derived from a combination of spin, valley, and orbital degrees of freedom\cite{novoselov2006unconventional, Falko2006,McCannKoshino},
in addition to cyclotron-orbit-center degeneracy. 
The orbital degree of freedom is present because quantized cyclotron orbits with two different form 
factors are nearly degenerate in this geometry, $n=0$ LLs that are maximally localized and
identical to the lowest orbital LLs in semiconductor quantum wells, 
and $n=1$ LLs that are 
less localized and similar to the first excited orbital LLs in semiconductor quantum wells. 
The many-body states formed within this eight-Landau-level space can be controlled by 
varying Zeeman field to favor a particular spin-polarizaiton, and by varying the displacement field 
between layers to favor a particular valley-polarization. 

\begin{figure}[H]
\begin{center}
\includegraphics[width=\linewidth]{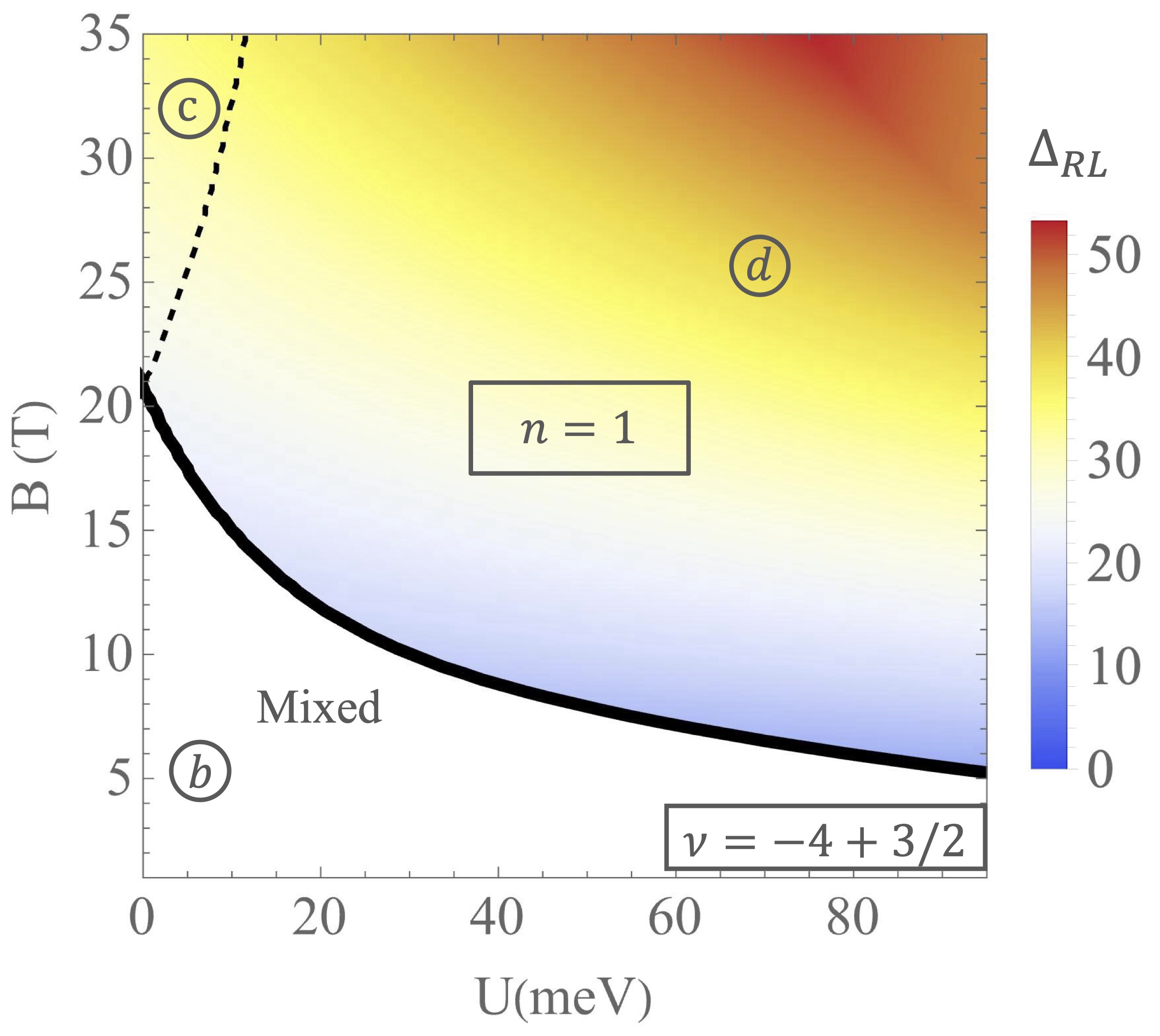}
\end{center}
\caption{Gap to remote levels $\Delta_{RL}$ in meV {\it vs.} 
orbital magnetic field $B$ and screened potential difference between layers $U$ 
when the $n=1$ LL is half-filled at $\nu= -4+ 3/2$, the filling factor we identify as 
being most favorable for isolated $n=1$ FQH physics.  This plot is for relative dielectric constant 
$\epsilon_r=5$, and Zeeman field $B_\mathrm{ZM}=35$ T.  
In mean-field-theory there is a phase-transition at weak fields 
to a state in which the valence LL has mixed $n=0$ and $n=1$ character
(white in the phase diagram).  
The boundary between mixed and $n=1$ states moves to weaker fields with $U$ 
in the illustrated low energy valley ($\nu <0$) case and to stronger fields in the high energy valley ($\nu>0$) case indicated by the dashed phase boundary line.  The circles mark phase boundary positions
discussed in greater detail below.
}
\label{Largest_Gap}
\end{figure}

In this Letter we address the subtle physics that
controls the orbital content of the many-body states.
This issue is of great importance to quantum Hall physics because of the interesting non-Abelian \cite{moore1991nonabelions} even-denominator states that form when $n=1$ LLs are fractionally occupied\cite{willett1987observation,Dean_2017,Young_2017_composite,Zhu_2022,Young_Gaps,Yazdani_Gap},
which have potential \cite{kitaev2003fault,sarma2005topologically, nayak2008non} quantum computing applications.  Indeed recent observations \cite{Young_Gaps,Yazdani_Gap} of large even-denominator 
gaps can occur in bilayer graphene.  Our goal is to shed light on how these states can be 
optimized.  We find that the competition between 
$n=1$ and $n=0$ orbitals is sensitive to the particle-hole symmetry breaking present in bilayer graphene
and to exchange interactions between the octet and the bilayer's sea of negative energy LLs,
as we explain below.  Our main conclusions are summarized in Fig.~\ref{Largest_Gap} where we plot estimates of the gap to 
remote LLs when the $n=1$ orbital is half-occupied.  As indicated in this figure we predict that the 
most favorable filling factor for isolated $n=1$ fractional quantum Hall (FQH) physics is $\nu=-4+3/2$,
and that the gap to remote levels is largest along a ridge in the two-dimensional displacement-potential ($U$) 
magnetic-field ($B$) control space.  Our calculations also predict that many-body states in which $n=0$ and 
$n=1$ are mixed will become prominent at weaker magnetic fields, possibly explaining an unexpected appearance of 
stripe charge-density-wave states in a recent STM study \cite{yazdani2023direct} of bilayer graphene FQH states.
These conclusions depend in detail on the way in which $U$ splits 
$n=0$ and $n=1$ orbitals in the bilayer's $K$ and $K'$ valleys, and on exchange-interactions between the 
octet LLs and the negative energy sea.  In mean-field theory the mixed $n=0$ and 
$n=1$ state (white region) has broken rotational symmetry and coherence between $n=0$ and $n=1$.

\textit{Non-interacting Model-}
We start from the established \cite{McCannKoshino,Jeil} low-energy two-band model
Hamiltonian of Bernal bilayer graphene in a perpendicular magnetic field $\bm B = - B\hat{z}$, 
neglecting trigonal warping but retaining particle-hole symmetry breaking:
\begin{align}
\begin{split}
& \hat{H}=\hat{h}_0+\hat{h}^\prime,  \hat{h}^\prime=\hat{h}_4+\hat{h}_\Delta, \\
& \hat{h}_0=-\frac{v^2}{t_1}\left(\begin{array}{cc} 0
& \left(\pi^{\dagger}\right)^2 \\
\pi^2 & 0
\end{array}\right) , \\
& \hat{h}^\prime=(\frac{2 v v_4}{t_1}+\frac{\Delta^{\prime} v^2}{t_1^{ 2}} )\left(\begin{array}{cc}
\pi^{\dagger} \pi & 0 \\
0 & \pi \pi^{\dagger}
\end{array}\right) 
\end{split}
\label{hamiltonian}
\end{align}
where $\pi = \xi (\hbar k_x + eA_x) + i(\hbar k_y + eA_y)$ is the complex kinetic momentum, $\xi=\pm1$ for $K/K^\prime$ valleys ,
the velocities are $v=-t_0 \sqrt{3} a/2 \hbar$ and $v_4= t_4 \sqrt{3} a/2 \hbar$, where $a=2.46$ \AA. 
The model parameters in Eq.~\ref{hamiltonian} ($t_0=-\gamma_0=-2.61$ eV, $t_1=\gamma_1=0.361$ eV, $t_4=\gamma_4=0.138$ eV, $\Delta^\prime=0.015$ eV) have been estimated from {\it ab initio} DFT calculations \cite{Jeil} and the matrix Hamiltonian acts on layer spinors.  (The $\gamma_i$'s correspond to the SWM \cite{SWM,SWM2} parameterization scheme.) 
The corresponding LL ladder operators are
$\tilde{\pi}=\frac{l_B}{\sqrt{2} \hbar} \pi$ and $\tilde{\pi}^{\dagger}$, where the magnetic length $l_B=\sqrt{\hbar/eB}$ 
and $[\tilde{\pi}, \tilde{\pi}^\dagger]=\xi$.

The Hamiltonian can be block diagonalized by noting that LL $n-2$ in one layer is coupled only to 
LL $n$ in the other layer.  The corresponding Hamiltonian blocks are:
\begin{equation}
    H= \begin{pmatrix} 
\tilde{\gamma}^\prime n & \tilde{\gamma}_0 \sqrt{(n-1)n}  \\
\tilde{\gamma}_0\sqrt{(n-1)n} & \tilde{\gamma}^\prime (n-1)
\end{pmatrix}
\end{equation}
where $\tilde{\gamma}^\prime=3 (\frac{a}{l_B})^2 \frac{t_0}{t_1} (\frac{1}{2} \frac{t_0}{t_1} \Delta^\prime-t_4) $
and $\tilde{\gamma}_0=-\frac{3}{2} (\frac{a}{l_B})^2 (\frac{t_0^2}{t_1})$. This layer Hamiltonian can be expanded in 
terms of layer Pauli matrices $H=b_i \sigma_i$ with $b_0^{n}= \tilde{\gamma}^\prime (2n-1)/2$, $b_1^{n}=\tilde{\gamma}_0 \sqrt{(n-1)n}$,
$b_2^{n}=0$ and $b_3^{n}= \tilde{\gamma}^\prime/2$, which allows us to write the eigenstates for the valence band for $n \ge 2$ as: $\ket{\Psi_n^-}=( \mathrm{sin} (\theta^{(n)}/2 )\ket{n}, - \mathrm{cos} (\theta^{(n)}/2 )\ket{n-2})$, where $\mathrm{cos}(\theta^{(n)})=b_3^{n}/\sqrt{(b_3^{n})^2+(b_1^{n})^2}$.  
There are in addition two zero energy modes $\ket{\Psi_0}=\left( \ket{0}, 0 \right)$ and $\ket{\Psi_1}=( \ket{1}, 0 )$,
which are degenerate in the particle-hole symmetric $\tilde{\gamma}^\prime=0$ limit.   Including $\tilde{\gamma}^\prime$ lifts this degeneracy; increasing the energy of $\ket{\Psi_1}$ relative to $\ket{\Psi_0}$ by $\tilde{\gamma}^\prime$.

\textit{Interactions with the Negative energy Sea-} Provided that we can neglect fluctuations of particle numbers in
LL's we can project the Hamiltonian to the octet,
provided that we add exchange self-energies due to interactions with the negative energy sea - the
non-relativistic quamtum mechanics limit of Lamb shifts.  
The density-matrix of the negative energy sea $ \hat{\rho}=\sum_{m,n=2}^\infty  \ket{\Psi^-_{n,m}} \bra{\Psi^-_{n,m}}$, where 
$m$ is an intra-Landau-level state label, produces a Fock field in the octet that is independent of guiding center,
but splits $n=0$ and $n=1$ by a Lamb shift 
$\Delta^X\equiv \Sigma_0-\Sigma_1 = \Delta^X_0+\delta \Delta^X$, where $\Delta^X_0=\sqrt{\pi/2}e^2/8 \epsilon_r l_B$ is the splitting in the particle-hole symmetric
$\tilde{\gamma}^\prime=0$ limit \cite{Shizuya,supmat} and $\epsilon_r$ is the dielectric constant.
The correction to the Lamb shift away from the particle-hole symmetric limit is
\begin{align}
\begin{split}
    \delta \Delta^X= &\frac{ e^2}{ \sqrt{2} \epsilon_r l_B}\sum_{n=1}^{\infty} \frac{1}{n !} \Gamma(n+\frac{1}{2}) \\&\times \left[(2n+1) c_n -(n+1) c_{n+1} - n c_{n-1}  \right] .
\end{split}
\label{Deltadelta}
\end{align}
where $\Gamma(x)$ is the Gamma function, $c_n=\cos{(\theta^{(n)})} /2$ 
for $n\ge2$ and zero otherwise. The Lamb shifts are identical in the two valleys, as required by the inversion symmetry of bilayer graphene. 
This is no longer true in presence of a displacement field, as discussed in the following section.  
\begin{figure}[H]
\begin{center}
\includegraphics[width=\linewidth]{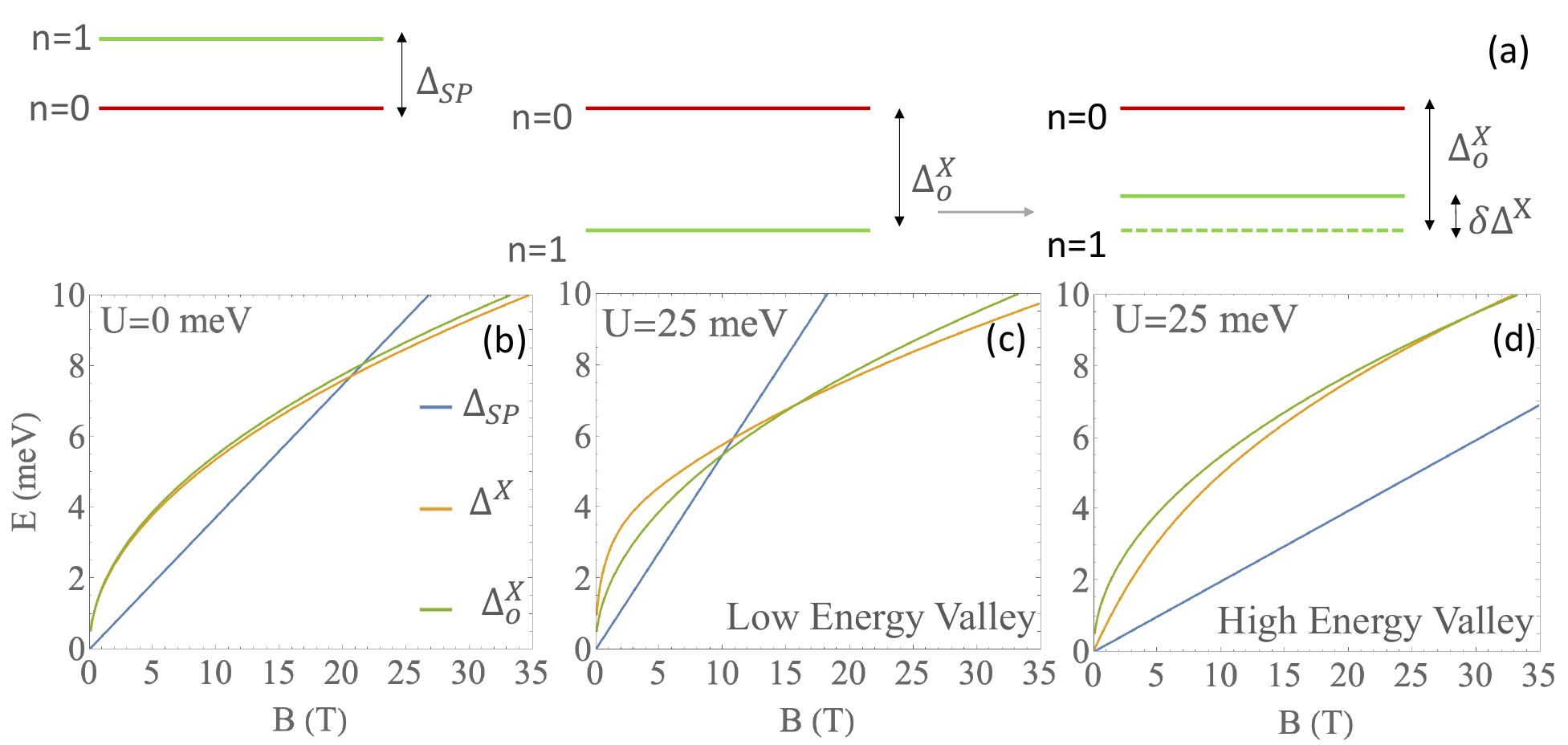}
\end{center}
\caption{a) Three effects that split the $n=0$ and $n=1$ levels, single-particle splitting, Lamb shifts in the particle-hole 
symmetric limit, and particle-hole assymmetry corrections to the Lamb shift.
b-d) Calculated values for single particle splitting $\Delta_{SP}$,
which favors initial occupation of $n=0$, and the Lamb shift $\Delta^{X}$ which favors initial occupation of $n=1$. 
These curves were calculated with dielectric constant
$\epsilon_r=5$ and displacement-fields potentials $U=0$ and $|U|= 25$ meV.}
\label{competition}
\end{figure}

Fig. \ref{competition} shows the magnitude of the terms contributing to the orbital splitting between the $n=0$ and $n=1$ LLs. The single particle splitting grows linearly with magnetic field and favors initial occupation of $n=0$ as the orbital levels are 
being filled. The splitting due to exchange interactions with the negative energy sea, which scales as $\sqrt{B}$ (the exchange energy scale is $ e^2/l_B$) and favors initial occupation of $n=1$, dominates at weak $B$.
For this reason $n=1$ is favored over $n=0$ at weak fields whereas $n=0$ is favored at stronger magnetic fields.
Because of this competition, the magnetic field $B_c$ at which $\Delta = \Delta_{SP} - \Delta^{X}$ crosses zero is an 
important characteristic magnetic field for bilayer FQH physics.  A clear preference for initial occupation of one of the 
two orbitals will result in sequential occupation and open up areas within the phase diagram with $n=1$ FQH physics.

\begin{figure}
\begin{center}
\includegraphics[width=\linewidth]{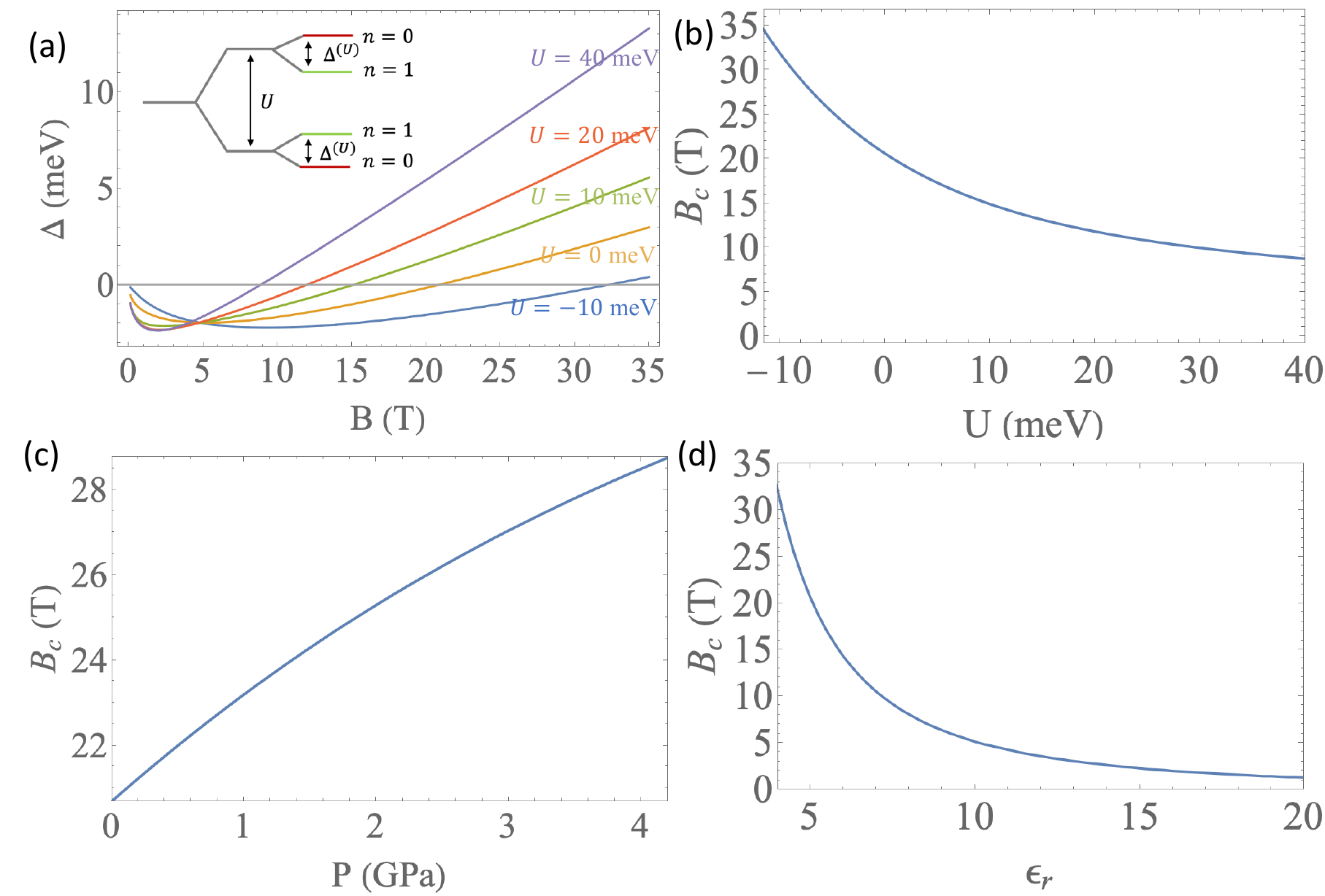}
\end{center}
\caption{Experimental control parameters that influence the splitting $\Delta$ between the $n=1$ and $n=0$ LLs by 
changing the critical magnetic field $B_c$ at which $n=0$ becomes lower in energy than $n=1$.
(a) Total splitting $\Delta$ in the low energy valley {\it vs.} magnetic field $B$ at different displacement potentials $U$.
The level schematic summarizes the valley dependence of the orbital shifts.  Note that the sign of $\Delta$ 
changes from negative to positive at a critical magnetic field $B_c$. 
(b) $B_c$ {\it vs.} $U$.  at pressure $P=0$ and dielectric constant $\epsilon_r=5$.  
(c) Dependence on pressure at $U=0, \epsilon_r=5$. (d) Dependence on the dielectric constant at $U=0, P=0$.  }
%\caption{(a) Difference between single particle splitting and exchange energy splitting for $\epsilon_r=6$ in the $K$ valley for different interlayer bias(b) Magnetic field at which the single particle splitting wins over Dirac Sea Exchange splitting as a function of the interlayer bias.}
\label{knobs}
\end{figure}

\textit{Experimental Control Parameters-} Dual gates allow the external potential difference between layers $U$ to be 
controlled by a displacement field, adding an extra single-particle term\cite{McCannKoshino} to the Hamiltonian \eqref{hamiltonian}:
\begin{equation}
\hat{h}_U = -\frac{U}{2}
\left(
\begin{array}{cc}
1 & 0 \\
0 & -1
\end{array}
\right)
+ \left(\frac{v}{\gamma_1}\right)^2 U
\left(
\begin{array}{cc}
\pi^{\dagger} \pi & 0 \\
0 & -\pi \pi^{\dagger}
\end{array}
\right)
\label{U_K}
\end{equation}
The displacement field influences both the single particle splitting and the Lamb shift. 
Both contributions break valley symmetry, but valley interchange combined with $U \rightarrow -U$ remains a symmetry. 
$h_U$ splits the valleys by $U$ and within each valley splits $n=0$ and $n=1$ by 
$\Delta^{(U)}= \frac{3}{2} (\frac{a}{l_B})^2 (\frac{t_0}{\gamma_1})^2 U$. 
$h_U$ also leads to a change of the layer polarization of the negative energy sea, changing the Lamb shift,
as illustrated in Fig. \ref{competition}(c,d). 
Importantly, as illustrated in the inset of Fig. \ref{knobs} (a), 
$\Delta^{(U)}$ raises the energy of $n=1$ compared to $n=0$ in 
the low energy valley and decreases it in the high energy valley. 
The field-dependence of the splitting between $n=0$ and $n=1$ in the low and 
high energy valleys at $|U|= 25$meV when both Lamb and single-particle 
splitting effects are included is illustrated in Figs. \ref{competition} (c,d).
The inset in Fig. \ref{knobs}(a) summarizes the effects that contribute to the energy difference
$\Delta$ between $n=0$ and $n=1$.  At weak fields $\Delta$ is negative because the 
Lamb shift effect dominates.  When $\Delta > 0$
orbital occupancy follows a $n=0$ then $n=1$ rule discussed further below.  The dependence on $U$ of the magnetic field $B_c$ at which $\Delta$ crosses zero is illustrated in \ref{knobs}b. 

The dependence of $\Delta_{SP}$ and $\delta \Delta^X$ \eqref{Deltadelta} on model parameters \cite{supmat},
%We phenomenologically model the effect of pressure as an increase of the interlayer hopping parameters $t_1, t_4$ such that $t_1(P)=t_1(1+\alpha P)$, $t_4(P)=t_4(1+\beta P)$  and assume $ \alpha= \beta$ (check. jusification?). The dependence of the two terms on the tight binding parameters is given by cite{Appendix}:
\begin{align}
\begin{split}
\Delta_{SP}&=\tilde{\gamma}^\prime=3 (\frac{a}{l_B})^2 \frac{t_0^2}{t_1} (\frac{1}{2} \frac{\Delta^\prime}{t_1} -\frac{t_4}{t_0})  \\
\delta \Delta^X &\sim \frac{ e^2}{ \epsilon_r l_B} |\frac{t_4}{t_0}-\frac{\Delta^\prime}{2 t_1}|   \sim \frac{\gamma^\prime}{\gamma_0},
\end{split}
\label{parametric_dependence}
\end{align}
suggests that they can be tuned by applying pressure \cite{MATBG_Pressure, carr2018pressure,jeil2018pressure}
to alter interlayer tunneling strengths.
%If we assume that $t_1(P)=t_1(1+\alpha P)$, $t_4(P)=t_4(1+\beta P)$  and $\alpha= \beta$
In the case of magic angle twisted bilayer graphene, for example, 
the shift of magic angle under pressure from $1.1 \degree$ to $1.27 \degree$ at a pressure of $1.3 $ GPa implies
that interlayer tunneling increases by $\alpha \approx 12 \% \, \mathrm{GPa}^{-1}$.
If both $t_1$ and $t_4$ change at the same rate with pressure we find from 
Eq. \eqref{parametric_dependence} that the largest influence on the orbital 
competition comes from a reduction in the single particle splitting by $ \approx 7 \%  \mathrm{GPa}^{-1}$.
The critical magnetic field at which the splitting flips signs is therefore 
expected to increase as we apply pressure, as illustrated in Fig. \ref{knobs} (c).
The competition between the single particle splitting and the Lamb shift is also sensitive to the dielectric environment,
as illustrated in Fig. \ref{knobs} (d).
Dielectric screening weakens the Lamb shift compared to the single particle splitting, 
thus reducing the magnetic field at which the 
$n=0$ orbital is occupied first. 

\begin{figure*}
\includegraphics[width=\linewidth]{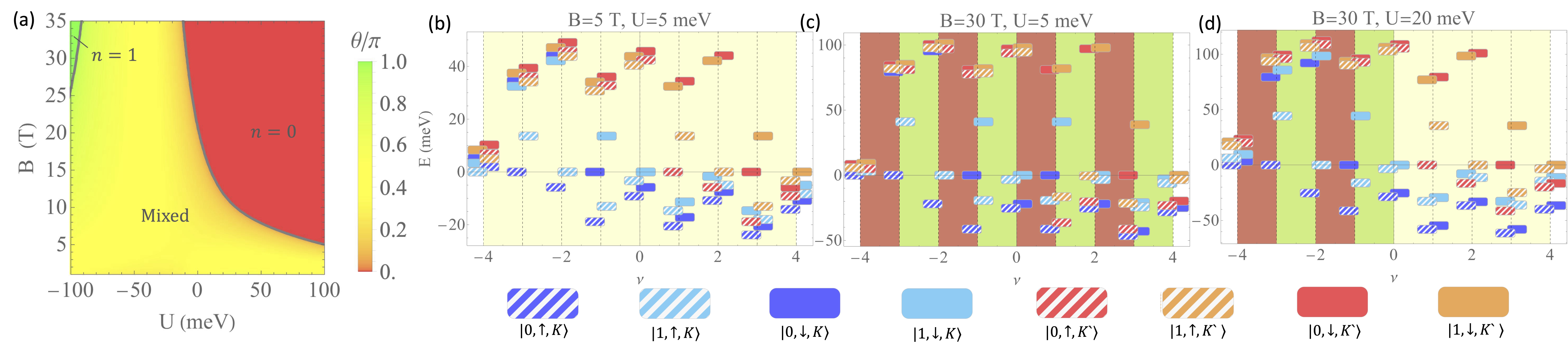}
        \caption{(a) Orbital polar angle $\theta/\pi$ in valley $K$ for dielectric constant $\epsilon_r=5$.
        ($\theta=0^{\circ}$ corresponds to a state that is fully $n=0$ polarized when the orbital space is half-filled, $\theta =\pi$ to full $n=1$ polarization and $\theta \in (0,\pi)$ to mixed orbital states.)
        In our convention valley $K$ is the low-energy valley for $U>0$ and the high energy valley for $U<0$.
        (b-d) Phase boundaries for the the three different regmies identified in Fig.~\ref{Largest_Gap}. The vertical positions of the boxes labelled by orbital, spin, and valley, indicate restricted ({\it i.e.} 
        calculated without coherence between orbitals) Hartree-Fock ground state quasiparticle energies at integer filling 
        factors .  The integer intervals are colored red (for $n=0$ FQH physics) 
        green (for $n=1$ FQH physics) and yellow for orbitally mixed states according to the calculated orbital polar angles.
        In most cases this assignment is consistent with the orbital of the lowest energy unoccupied orbital at the left
        side of the integer interval and the orbital of the highest occupied orbital (chosen as the zero of energy)
        at the right hand side of the interval.
        For mixed states, the quasiparticle orbital can change from $n=1$ to $n=0$ as the integer interval is filled. }
    \label{octet_plots}
\end{figure*}
\textit{The Flavorless Limit-} We now discuss the physical implications of the orbital splittings discussed above, starting with the 
simplest case in which spin and valley states are occupied sequentially because of large spin and valley Zeeman effects.
We will adopt the convention that the filling factor $\nu=-1$ when both orbitals are empty, $\nu=0$ when the orbital 
space is half-filled and $\nu=1$ when the orbital space is fully occupied.
The one-body shifts directly determine which orbital is 
first occupied at $\nu$=-1$^+$.  As the orbital space is filled, interactions within the orbital space also play a role.
To judge whether one of the two orbitals is fully occupied at $\nu=0$, we consider a Slater determinant state 
with coherence between orbitals,
$\ket{\Psi}= \cos{(\theta/2})\ket{0} + e^{i \phi} \sin{(\theta/2)}\ket{1}$, 
for each state within a LL.  In this state the $n=0$ orbitals are fully occupied 
when the variational parameter $\theta=0$ and the $n=1$ state is fully occupied at $\theta=\pi$.
Because the number of electrons within each orbital is not a good quantum number these states 
are not exact eigenstates of the many-electron Hamiltonian even at $\theta=0,\pi$, although we will take the 
view that they are accurate in these limits.  The mixed-orbital states at intermediate $\theta$ break 
rotational symmetry.  The possibility that these inter-orbital coherent states might be ground states at some
points in the phase diagram is interesting, but we recognize that this order might not survive quantum fluctuations. The energy of this variational state is \cite{supmat}

\begin{equation}
    \frac{E}{N_\phi}=-\frac{1}{64}\sqrt{\frac{\pi}{2}} \frac{ e^2}{ \epsilon_r l_B} \left (4 \cos{(\theta)}-\cos{(2\theta)}\right) + \Delta \sin^2(\theta/2),
    \label{eq:Ecoherent}
\end{equation}
where $N_\phi$ is the LL degeneracy. The first term on the right hand side of 
Eq.~\ref{eq:Ecoherent} is an exchange energy that favors $n=0$, while the second term captures
the one-body splittings and favors $n=0$ for $\Delta>0$ $(B>B_c)$ and $n=1$ for $\Delta <0$ $(B<B_c)$.
For $\Delta>0$ both terms favor $n=0$ and the energy is minimized at $\theta=0$. 
When the interaction term is included, there is a range of negative $\Delta$ over which 
mixed states with intermediate $ \theta \in (0,\pi) $ minimize the energy.  

Fig. \ref{octet_plots} (a) summarizes the phase diagram {\it vs.} $U$ and $B$.
At $U=0$, $\theta$ decreases continuously through intermediate values,
eventually reaching zero at $B_c$.  In the low-energy valley $U$
increases the energy of the $n=1$ orbital relative to $n=0$ and therefore expands
the $n=0$ phase stability to weaker magnetic fields.  In the high-energy valley, 
$U$ lowers $n=1$ relative to $n=0$.  States in which $n=1$ is 
occupied before $n=0$ are reached only at extremely large $U$ and $B$.
To fully understand the implications of these results for the filling factor dependence of 
orbital competition in bilayer graphene, the valley and spin degrees of freedom must be restored.

\textit{Adding Spin and Valley-} 
Non-abelian quasiparticles are, as far as we are aware, a characteristic of 
correlations in spin and valley polarized partially occupied $n=1$ orbitals that 
are not mixed with $n=0$ orbitals.  Assuming that spins and valleys are 
occupied sequentially and that $n=0$ orbitals have lower energy than
$n=1$ orbitals (see below), the favorable filling factor ranges for $n=1$ FQH 
physics are $i < \nu < i+1$, and the favorable filling factors of 
the $5/2$-like incompressible states is $i+1/2$ where $i$ is an odd integer.

The competition between $n=1$ and $n=0$ orbitals is complicated by 
spin and valley degrees of freedom, which can be manipulated by in-plane magnetic fields
that allow the Zeeman coupling to be tuned independently of the orbital magnetic field,
and by displacement fields.  The displacement potential $U$ can be viewed as 
a valley Zeeman coupling because it splits the energy of the two valleys 
as discussed above.
We consider the limit of relatively large Zeeman, and valley Zeeman such that we occupy
the four spin/valley flavors consecutively in energetic order.  
Except at very small $|U|$ the largest energy scale is the valley Zeeman splitting set by the displacement potential.  Thus, as the system is doped from $\nu=-4$ to $\nu=0$, we are normally filling one valley. 
Due to the Zeeman splitting, we first fill one spin (majority spin) from $\nu=-4$ to $\nu=-2 $ then the other (minority) spin from $\nu=-2 $ to $\nu=0$. We then fill other valley in the same fashion.  The key issue is 
whether in each flavor sector $n=0$ and $n=1$ orbitals are occupied sequentially or simultaneously.

For relatively weak magnetic fields and $|U|$ (See Fig. \ref{octet_plots} b and point $b$ in Fig.~\ref{Largest_Gap}), 
our analysis predicts orbitally mixed states, unfavorable for $n=1$ FQH physics at all filling factors.
In this regime the Lamb shift dominates giving $\Delta <0$, but not sufficiently to overcome exchange within the 
active LL.  At strong magnetic field, but weak displacement potential (Fig. \ref{octet_plots} c and 
point $c$ in Fig.~\ref{Largest_Gap}) the single particle contribution dominates 
favoring $n=0$ over $n=1$, and we always consecutively fill $n=0$ then $n=1$.  The sizes of the gaps between
$n=0$ and $n=1$ (see below) are however larger for $\nu<0$ than for $\nu>0$ because $|U|$ lowers the $n=1$ energy in the high energy valley.  At strong magnetic field and strong displacement potential ( Fig. \ref{octet_plots} d
and point $d$ in Fig.~\ref{Largest_Gap}) the valley-dependent orbital splitting is larger and 
produces a profound electron-hole asymmetry where we get consecutive occupation of $n=0$ and $n=1$ for $\nu<0$, and mixed states 
unfavorable for $n=1$ FQH physics for $\nu>0$. 

\begin{figure}
\begin{center}
\includegraphics[width= \linewidth]{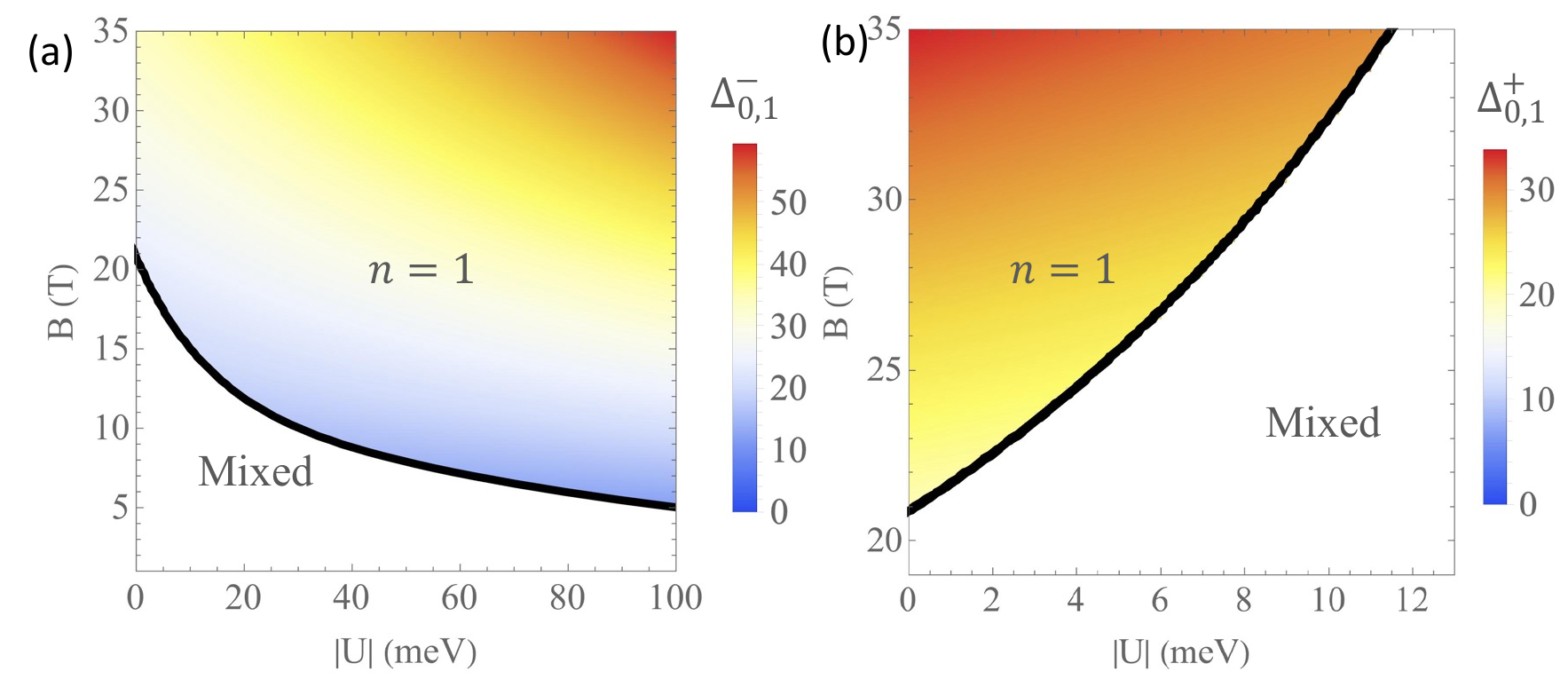}
\end{center}
\caption{Gap between the $n=1$ and $n=0$ LLs with the $n=0$ LL fully occupied and the $n=1$ LL half-filled (a) in the low energy valley and (b) in the high energy valley. White regions mask the parameter regime in which the ground state is expected to be 
orbitally mixed.}
\label{Sector_Gap}
\end{figure}

\textit{Optimal Conditions for non-Abelian quasiparticles -} Given our understanding of the energetic competition between
orbitals, we can assess when pure $n=1$ correlations are most robust against quantum fluctuations
by calculating the energy required to add an electron to a half-filled $n=1$ orbital and remove it from a full lower energy $n=0$ orbital.
Quantum fluctuations are expected, for example, to be important 
when the remote-level gap is small and to reduce the gaps of non-Abelian  incompressible 
states\cite{PRX_2015_Mixing,luo2016missing,KunYang_Review}. If the $n=0$ orbital is full, the removal energy can be calculated exactly but the addition energy must be approximated.
In Figs. \ref{Sector_Gap}(a,b) we show gaps calculated using mean-field Hartree-Fock estimates. 
The gap is largest in the low energy valley at larger magnetic fields and displacement potentials, and largest in the high energy value at larger magnetic fields and weaker displacement potentials.
Only in the low-energy valley, normally occupied for $\nu < 0$, can the gap be increased by applying displacement fields.
Robustness against quantum fluctuations and against disorder in termperature requires that both the 
gaps in Fig.~\ref{Sector_Gap} and in Fig.~\ref{Largest_Gap} be large.
Displacement fields are desirable for isolating $n=1$ LLs because they increase the orbital 
separation in the low-energy valleys - identifying negative filling factors as the attractive regime.
At $\nu=-4+3/2$ the $n=1$ orbital of the majority-valley and majority-spin LL is half-filled in much of the 
phase diagram and is well separated in energy from other orbitals in both directions
(See Fig.~\ref{octet_plots} (b)-(d)). $\nu=-4+3/2$, where $n=1$ majority-valley majority
-spin is half-filled in 
much of the phase diagram, is preferred over $\nu=-2+3/2$ because $n=1$ majority-valley majority-spin LLs lie just below the Fermi level in the latter case.  We conclude therefore that   
the purest $n=1$ FQH physics will occur at $\nu=-4+3/2$ highlighted in Fig.~\ref{Largest_Gap}.
Below $U \sim 80$ meV, the closest level is then to the full $n=0$ level in same spin/valley sector, 
and therefore also increases with increasing magnetic field.  For large enough displacement field, 
the $n=1$ LL is raised enough in energy that it approaches the $n=0$ LL in the closest spin/valley sector,
and thus the gap starts decreasing. That means that there will be an optimal value of 
displacement and magnetic fields for which the gap is maximized.

\textit{Discussion-} Some experimental studies \cite{Yacobi_2014,Dean_2017,Young_2017,Young_Gaps} of small $U$ bilayer graphene FQH 
physics have found that the pattern of alternate filling of $n=0$ and $n=1$ seen in Fig. \ref{octet_plots} (c)
is established at weaker fields than we have estimated theoretically.  We ascribe this quantitative
discrepancy to the absence in our calculations of Coulomb screening effects due to mixing between the octet and 
higher energy LLs.  Weaker interactions produced by that screening will reduce the size of the 
Lamb shift effect, which favors orbitally mixed states, and move the transition 
field to smaller values of $B$ at every displacement field $U$.   We therefore expect that a 
more accurate treatment of screening 
would yield a phase diagram in Fig. \ref{octet_plots} (a) that is similar in shape 
but has a smaller magnetic field scale, just as enhanced dielectric screening 
is responsible for the reduction in transition fields illustrated in Fig. \ref{knobs} (d).

We have mainly emphasized the goal of finding pure $n=1$ FQH physics within the bilayer graphene 
octet.  The orbitally mixed states which we expect at weaker magnetic fields, could however 
hold their own surprises.  It has, for example, been predicted in earlier work that 
orbitally mixed states are likely to be unstable to long-wavelength 
translational symmetry breaking \cite{cote2010orbital}. States with coherence between orbitals, which we 
have examined within mean-field theory, will break rotational symmetry if they survive quantum
fluctuations.  These properties could be connected to recent observations in bilayer graphene 
\cite{yazdani2023direct} of Wigner crystal
and rotational symmetry broken states at weak magnetic fields-though at fractional filling factor. 

The competition between orbitals that we study 
here in bilayer graphene is also expected to play a role in the strong magnetic field 
physics of bulk graphite and rhombohedral graphene multilayers. 
The behavior of bulk graphite in a strong magnetic field \cite{esquinazi2016book,wang2020critical,zhu2019graphite,marcenat2021wide} is a long-standing puzzle.
In a strong field, $n$ layer rhombohedral graphene has $n$ nearly degenerate LLs \cite{min2008chiral}, whose 
relative energies will also be influenced by both single-particle and Lamb shift effects.  
The FQH physics of this case has received recent theoretical attention in which 
non-Abelian states with Ising and Fibonacci anyons were proposed \cite{xiaogang2023non}.
This rich generalization of the bilayer case deserves future attention from both 
theory and experiment. 

\textit{Acknowledgments-} This work was supported by Welch Foundation grant TBF1473 and by the 
Simons Foundation.  

\bibliography{refs.bib}

\newpage

\onecolumngrid 
\appendix
\section{Supplemental Material}
\subsection{Negative Energy Sea Exchange Interactions}
To derive the expression for the Lamb shift , we start with the density matrix for the valence band at zero temperature:
\begin{equation}
    \hat{\rho}=\sum_{\substack{X, n=2}}^\infty  \ket{\Psi_{n,X}} \bra{\Psi_{n,X}}
\end{equation}
where $n$ is the LL index and $X$ is an intra-Landau-level guiding center label. The density matrix in layer x LL basis reads:
\begin{equation}
\rho_{n X, n^\prime, X'}=
    \begin{pmatrix}
        \sin^2({\frac{\theta^{(n)}}{2}}) \delta_{n,n^\prime} \Theta(n-2) &  -\cos({\frac{\theta^{(n)}}{2}}) \sin({\frac{\theta^{(n)}}{2}}) \delta_{n^\prime,n-2} \Theta(n-2) \\
        -\cos({\frac{\theta^{(n^\prime)}}{2}}) \sin({\frac{\theta^{(n^\prime)}}{2}}) \delta_{n,n^\prime-2} \Theta(n) & \cos^2({\frac{\theta^{(n+2)}}{2}}) \delta_{n,n^\prime} \Theta(n)
    \end{pmatrix} \delta_{X, X'}
\label{DM}
\end{equation}

% \begin{equation}
% \rho_{n, n^\prime}=
%     \begin{pmatrix}
%         \cos^2({\frac{\theta^{(n)}}{2}}) \delta_{n,n^\prime} \Theta(n-2) &  \textcolor{red}{\pm*}\cos({\frac{\theta^{(n)}}{2}}) \sin({\frac{\theta^{(n)}}{2}}) \delta_{n^\prime,n-2} \Theta(n-2) \\
%         \textcolor{red}{\pm*}\cos({\frac{\theta^{(n^\prime)}}{2}}) \sin({\frac{\theta^{(n^\prime)}}{2}}) \delta_{n,n^\prime-2} \Theta(n) & \sin^2({\frac{\theta^{(n)}}{2}}) \delta_{n,n^\prime} \Theta(n)
%     \end{pmatrix}
% \label{DM}
% \end{equation}

We now write our states as $\ket{A, n, X}$ where $A$ is a layer label, so the coulomb matrix elements read:
\begin{align}
    \langle A_1^\prime n_1^\prime X_1^\prime; A_2^\prime n_2^\prime X_2^\prime|V|A_1 n_1 X_1; A_2 n_2 X_2 \rangle &= \frac{1}{2 \pi} \int d\mathbf{q} \frac{\kappa e^2}{q} \langle A_1^\prime n_1^\prime X_1^\prime; A_2^\prime n_2^\prime X_2^\prime|e^{i \mathbf{q}.\mathbf{r}}|A_1 n_1 X_1; A_2 n_2 X_2 \rangle \\
    & = \delta_{A_1^\prime, A_1} \delta_{A_2^\prime, A_2} \langle n_1^\prime X_1^\prime;  n_2^\prime X_2^\prime|e^{i \mathbf{q}.\mathbf{r}}|n_1 X_1; n_2 X_2 \rangle
\end{align}
where $\epsilon_r \kappa=1/4 \pi \epsilon_0=14.3996$ eV\AA $e^{-2}$ (in the main text we absorb $\kappa$ in the units of $e$), and \cite{Cote_MacDonald}:
\begin{align}
\begin{split}
    \langle n_1^\prime X_1^\prime;  n_2^\prime X_2^\prime|e^{i \mathbf{q}.\mathbf{r}}|n_1 X_1; n_2 X_2 \rangle=& F_{n_1^\prime, n_1}(q) F_{n_2, n_2^\prime}^*(q)e^{i q_x (X_1^\prime+X_1)/2} e^{-i q_x (X_2^\prime+X_2)/2} \\
    &\times \delta_{X_1^\prime, X_1+ l_B^2 q_y} \delta_{X_2, X_2^\prime+ l_B^2 q_y}
\end{split}
\end{align}
The exchange part of the Hamiltonian reads:
\begin{align}
\begin{split}
    H^{X}_{a,b}&= - \sum_{\substack{c,d}}\langle a;c |V| d;b\rangle \rho_{d,c} 
    \\
    &= - \sum_{\substack{\{\mu n X\}_c \\ \{\mu n X\}_d}} \langle \{\mu n X\}_a;\{\mu n X\}_c |V| \{\mu n X\}_d;\{\mu n X\}_b\rangle \rho_{\mu_d n_d, \mu_c n_c} \delta_{X_c,X_d} \\ 
    &= - \sum_{\substack{\{\mu n\}_c \\ \{\mu n \}_d \\ X_c}} \langle \{\mu n X\}_a;\{\mu n X\}_c |V| \{\mu n X\}_d;\{\mu n X\}_b\rangle \rho_{\mu_d n_d, \mu_c n_c}  \\ 
    &=-  \frac{1}{2 \pi} \int d\mathbf{q} \frac{\kappa e^2}{q}\sum_{\substack{\{\mu n\}_c  \\ \{\mu n \}_d \\ X_c}} \delta_{\mu_a, \mu_d} \delta_{\mu_c, \mu_b} F_{n_a, n_d}(q) F_{n_b, n_c}^*(q)e^{i q_x (X_a+X_c)/2} e^{-i q_x (X_b+X_c)/2} \\& \mspace{190mu} \times \delta_{X_a, X_c+ l_B^2 q_y} \delta_{X_b, X_c+ l_B^2 q_y} \rho_{\mu_d n_d, \mu_c n_c}
\end{split}
\end{align}
Summing over $X_c$:
\begin{align}
    H^{X}_{a,b}&= - \frac{1}{2 \pi} \int d\mathbf{q} \frac{ \kappa e^2}{q} \sum_{\substack{\{\mu n\}_c \\ \{\mu n \}_d}} \delta_{\mu_a, \mu_d} \delta_{\mu_c, \mu_b} F_{n_a, n_d}(q) F_{n_b, n_c}^*(q) \delta_{X_a, X_b} \rho_{\mu_d n_d, \mu_c n_c}
\end{align}
Since the Hamiltonian is diagonal in $X$, we drop the $\delta_{X_a, X_b}$, keeping in mind this is for one intra-LL index. Summing over $\mu_c,\mu_d$:
\begin{align}
    H^{X}_{a,b}&= - \frac{1}{2 \pi} \int d\mathbf{q} \frac{\kappa e^2}{q} \sum_{n_c, n_d}F_{n_a, n_d}(q) F_{n_b, n_c}^*(q) \rho_{\mu_a n_d, \mu_b n_c}
\label{HX}
\end{align}

Substituting in $\rho$ from eq. \eqref{DM}:

\begin{align}
\begin{split}
    H^{X}_{n_a,n_b}&= - \frac{1}{2 \pi} \int d\mathbf{q} \frac{ \kappa e^2}{q} \sum_{n_c, n_d}F_{n_a, n_d}(q) F_{n_b, n_c}^*(q) \times \\ &\begin{pmatrix}
        \sin^2({\frac{\theta^{(n_d)}}{2}}) \delta_{n_c,n_d} \Theta(n_d-2) &  -\cos({\frac{\theta^{(n_d)}}{2}}) \sin({\frac{\theta^{(n_d)}}{2}}) \delta_{n_c,n_d-2} \Theta(n_d-2) \\
        -\cos({\frac{\theta^{(n_c)}}{2}}) \sin({\frac{\theta^{(n_c)}}{2}}) \delta_{n_d,n_c-2} \Theta(n_d) & \cos^2({\frac{\theta^{(n_d+2)}}{2}}) \delta_{n_c,n_d} \Theta(n_d)
    \end{pmatrix} \\
    % &=- \frac{1}{2 \pi} \int d\mathbf{q} \frac{\kappa e^2}{q} \sum_{n_c}F_{n_a, n_d}(q) F_{n_b, n_c}^*(q) \times \\ &\begin{pmatrix}
    %     \cos^2({\frac{\theta^{(n_c)}}{2}}) \delta_{n_c,n_d} \Theta(n_c-2) &  \textcolor{red}{\pm*}\cos({\frac{\theta^{(n_c)}}{2}}) \sin({\frac{\theta^{(n_c)}}{2}}) \delta_{n_d,n_c-2} \Theta(n_c-2) \\
    %     \textcolor{red}{\pm*}\cos({\frac{\theta^{(n_d)}}{2}}) \sin({\frac{\theta^{(n_d)}}{2}}) \delta_{n_c,n_d-2} \Theta(n_c) & \sin^2({\frac{\theta^{(n_c)}}{2}}) \delta_{n_c,n_d} \Theta(n_c)
    % \end{pmatrix} \\
\end{split}
\end{align}
Using the delta and step functions in evaluating the sums, we get:
\begin{align}
\begin{split}
    H^X_{A;n_a,A; n_b}&=- \frac{1}{2 \pi} \int d\mathbf{q} \frac{\kappa e^2}{q} \sum_{n_d=2}^\infty F_{n_a, n_d}(q) F_{n_b, n_d}^*(q) \sin^2({\frac{\theta^{(n_d)}}{2}})  \\
    H^X_{B;n_a,B; n_b}&=- \frac{1}{2 \pi} \int d\mathbf{q} \frac{\kappa e^2}{q} \sum_{n_d=0}^\infty F_{n_a, n_d}(q) F_{n_b, n_d}^*(q) \cos^2({\frac{\theta^{(n_d+2)}}{2}})  \\
    H^X_{A;n_a,B; n_b}&= \frac{1}{2 \pi} \int d\mathbf{q} \frac{\kappa e^2}{q} \sum_{n_d=2}^\infty F_{n_a, n_d}(q) F_{n_b, n_d-2}^*(q) \cos({\frac{\theta^{(n_d)}}{2}})  \sin({\frac{\theta^{(n_d)}}{2}})   \\
    H^X_{B;n_a,A; n_b}&= \frac{1}{2 \pi} \int d\mathbf{q} \frac{\kappa e^2}{q} \sum_{n_c=2}^\infty F_{n_a, n_c-2}(q) F_{n_b, n_c}^*(q)  \cos({\frac{\theta^{(n_c)}}{2}})  \sin({\frac{\theta^{(n_c)}}{2}})   \\
\end{split}
\label{exchange_ham}
\end{align}

Let us now focus on the $\mathbf{q}$ integral, the $\mathbf{q}$ dependence only comes from the Coulomb potential and the form factor:

\begin{align}
    \int d\mathbf{q} \frac{1}{q} F_{n_a, n_d}(\mathbf{q}) F_{n_b, n_c}^*(\mathbf{q})&=\int_0^\infty q dq \int_0^{2\pi} d\theta_q \frac{1}{q} F_{n_a, n_d}(\mathbf{q}) F_{n_b, n_c}^*(\mathbf{q}) \\
    &=\int_0^\infty  dq \int_0^{2\pi} d\theta_q F_{n_a, n_d}(\mathbf{q}) F_{n_b, n_c}^*(\mathbf{q})
\label{integral}
\end{align}
The explicit expression for the form factor is given by \cite{Cote_MacDonald}:
\begin{align}
F_{n^{\prime}, n}(\mathbf{q})&=\left(\frac{n !}{n^{\prime} !}\right)^{1 / 2}\left(\frac{\left(-q_y+i q_x\right) l_B}{\sqrt{2}}\right)^{n^{\prime}-n} \exp \left(\frac{-q^2 l_B^2}{4}\right) L_n^{n^{\prime}-n}\left(\frac{q^2 l_B^2}{2}\right) 
\end{align}
for $n \le n^\prime$. We can use $F_{n,n^\prime}(\mathbf{q})=F^*_{n^\prime,n}(-\mathbf{q})$ to get the expression for $n\ge n^\prime$:
\begin{align}
F_{n^{\prime}, n}(\mathbf{q})&=\left(\frac{n !}{n^{\prime} !}\right)^{1 / 2}\left(\frac{\left(q_y+i q_x\right) l_B}{\sqrt{2}}\right)^{n-n^{\prime}} \exp \left(\frac{-q^2 l_B^2}{4}\right) L_{n^\prime}^{n-n^{\prime}}\left(\frac{q^2 l_B^2}{2}\right) \\
&\sim e^{i\theta_q (n-n^\prime)}
\end{align}
This is, the only dependence on $\theta_q$, so the angular integral in \eqref{integral} gives (for $n_a<n_d, n_b <n_c $):
\begin{align}
    \int_0^{2\pi}d\theta_q e^{i\theta_q (n_d-n_a)} e^{-i\theta_q (n_c-n_b)}
\end{align}
Substituting this in the Hamiltonian gives:
\begin{align}
\begin{split}
    H^X_{A;n_a,A; n_b}& \sim \int_0^{2\pi} d\theta_q e^{i(n_c-n_a) \theta_q} e^{-i(n_c-n_b)\theta_q}=\int_0^{2\pi} d\theta_q e^{i(n_b-n_a)\theta_q}=2 \pi \delta_{n_a,n_b} \\
    H^X_{B;n_a,B; n_b}& \sim \int_0^{2\pi} d\theta_q e^{i(n_c-n_a)\theta_q} e^{-i(n_c-n_b)\theta_q}=\int_0^{2\pi} d\theta_q e^{i(n_b-n_a)\theta_q}=2 \pi \delta_{n_a,n_b} \\
    H^X_{A;n_a,B; n_b}& \sim \int_0^{2\pi} d\theta_q e^{i(n_c-n_a)\theta_q} e^{-i(n_c-n_b-2)\theta_q}=\int_0^{2\pi} d\theta_q e^{i(n_b+2-n_a)\theta_q}=2 \pi \delta_{n_a,n_b+2} \\
    H^X_{B;n_a,A; n_b}& \sim \int_0^{2\pi} d\theta_q e^{i(n_d-2-n_a)\theta_q} e^{-i(n_d-n_b)\theta_q}=\int_0^{2\pi} d\theta_q e^{i(n_b-n_a-2)\theta_q}=2 \pi \delta_{n_a,n_b-2} \\
\end{split}
\end{align}
So \eqref{exchange_ham} becomes:
\begin{align}
\begin{split}
    H^X_{A;n_a,A; n_b}&=- \kappa e^2 \int_0^\infty dq  \sum_{n_d=2}^\infty F_{n_a, n_d}(q) F_{n_b, n_d}^*(q) \sin^2({\frac{\theta^{(n_d)}}{2}}) \delta_{n_a,n_b} \\
    H^X_{B;n_a,B; n_b}&=- \kappa e^2 \int_0^\infty dq  \sum_{n_d=0}^\infty F_{n_a, n_d}(q) F_{n_b, n_d}^*(q) \sin^2({\frac{\theta^{(n_d+2)}}{2}}) \delta_{n_a,n_b} \\
    H^X_{A;n_a,B; n_b}&= \kappa e^2 \int_0^\infty dq  \sum_{n_d=2}^\infty F_{n_a, n_d}(q) F_{n_b, n_d-2}^*(q)  \cos({\frac{\theta^{(n_d)}}{2}})  \sin({\frac{\theta^{(n_d)}}{2}}) \delta_{n_a,n_b+2}  \\
    H^X_{B;n_a,A; n_b}&= \kappa e^2\int_0^\infty dq  \sum_{n_c=2}^\infty F_{n_a, n_c-2}(q) F_{n_b, n_c}^*(q) \cos({\frac{\theta^{(n_c)}}{2}})  \sin({\frac{\theta^{(n_c)}}{2}}) \delta_{n_a,n_b-2}  \\
\end{split}
\end{align}

The relevant sector is the AA sector for $n_a,n_b={0,1}$ which is diagonal and decoupled from the rest of the Hamiltonian. These read:
\begin{align}
\begin{split}
    H^X_{A;0,A; 0}\equiv \Sigma_0&=-  \kappa e^2 \int_0^\infty dq  \sum_{n=2}^\infty F_{0, n}(q) F_{0, n}^*(q) \sin^2({\frac{\theta^{(n)}}{2}})= -  \kappa e^2 \int_0^\infty dq  \sum_{n=2}^\infty e^{-q^2 l_B^2/2} |\tilde{F}_{0, n}(q)|^2 \sin^2({\frac{\theta^{(n)}}{2}}) \\
    H^X_{A;1,A; 1}\equiv \Sigma_1&=- \kappa e^2 \int_0^\infty dq  \sum_{n=2}^\infty F_{1, n}(q) F_{1, n}^*(q) \sin^2({\frac{\theta^{(n)}}{2}})=- \kappa e^2 \int_0^\infty dq  \sum_{n=2}^\infty  e^{-q^2 l_B^2/2}|\tilde{F}_{1, n}(q)| \sin^2({\frac{\theta^{(n)}}{2}}) 
\end{split}
\label{Sigmas0}
\end{align} 

where  $\tilde{F}_{n,n^\prime} \equiv e^{q^2 l_B^2/4} F_{n,n^\prime} $.
% so \eqref{Sigmas0} becomes:
% \begin{align}
%     \begin{split}
%         \Sigma_0&=-  \kappa e^2 \int_0^\infty dq  \sum_{n=2}^\infty e^{-q^2 l_B^2/2} |\tilde{F}_{0, n}(q)|^2 \sin^2({\frac{\theta^{(n)}}{2}}) \\
%          \Sigma_1&=- \kappa e^2 \int_0^\infty dq  \sum_{n=2}^\infty  e^{-q^2 l_B^2/2}|\tilde{F}_{1, n}(q)| \sin^2({\frac{\theta^{(n)}}{2}}) 
%     \end{split}
% \label{Sigmas1}
% \end{align}

In the particle hole symmetric limit $b_3=0$, $\mathrm{cos}(\theta^{(n)})=\frac{b_3^{n}}{\sqrt{(b_3^{n})^2+(b_1^{n})^2}}=0$, and $\cos{(\theta^{(n)}/2)}=\sqrt{\frac{1+\cos{(\theta^{(n)}})}{2}}=1/\sqrt{2}$ and $\sin{(\theta^{(n)}/2)}=\sqrt{\frac{1-\cos{(\theta^{(n)}})}{2}}=1/\sqrt{2}$ and \eqref{Sigmas0} becomes:

\begin{align}
    \begin{split}
         \Sigma_0&=- \frac{1}{2} \kappa e^2 \int_0^\infty dq  \sum_{n=2}^\infty e^{-q^2 l_B^2/2} |\tilde{F}_{0, n}(q)|^2 \\
        \Sigma_1&=- \frac{1}{2} \kappa e^2 \int_0^\infty dq  \sum_{n=2}^\infty  e^{-q^2 l_B^2/2}|\tilde{F}_{1, n}(q)| ^2
    \end{split}
\end{align}

The trick is then to use the identity \cite{Shizuya}:
\begin{equation}
    \sum_{n=0}^\infty |\tilde{F}_{i, n}(q)|= e^{q^2 l_b^2/2}
\end{equation}
for which we need to add and subtract the $n=0$ and $n=1$ terms so that we can write:
\begin{align}
    \begin{split}
         \Sigma_0&=- \frac{1}{2} \kappa e^2 \int_0^\infty dq   e^{-q^2 l_B^2/2}(e^{-q^2 l_B^2/2}- |\tilde{F}_{0, 0}(q)|^2-|\tilde{F}_{0, 1}(q)|^2) \\
         \Sigma_1&=- \frac{1}{2} \kappa e^2 \int_0^\infty dq   e^{-q^2 l_B^2/2}(e^{-q^2 l_B^2/2}- |\tilde{F}_{1, 0}(q)|^2-|\tilde{F}_{1, 1}(q)|^2) \\
    \end{split}
\end{align}
The first term in each is divergent. However, it is the same for both so it cancels out the difference is finite:

\begin{align}
    \Delta^X_0=\Sigma_0-\Sigma_1=\frac{1}{2} \kappa e^2 \int_0^\infty dq e^{-q^2 l_B^2/2}(|\tilde{F}_{0,0}(q)|^2-|\tilde{F}_{1,1}(q)|^2)= \frac{1}{8} \sqrt{\frac{\pi}{2}} \frac{\kappa e^2}{l_B}
\end{align}
where $\tilde{F}_{0,0}(q)=1$ and $\tilde{F}_{1,1}(q)=1-\frac{l_B^2 q^2}{2}$.

Without particle-hole symmetry, we cannot use the same trick because of the $\sin^2({\frac{\theta^{(n)}}{2}})$ factors in \eqref{Sigmas0}. Instead, we write:

\begin{align}
    \Delta^X=\Sigma_0-\Sigma_1 &=-\kappa e^2 \int_0^\infty dq e^{-q^2 l_B^2/2}\sum_{n=2}^\infty \sin^2{\frac{\theta^{(n)}}{2}}(|\tilde{F}_{0,n}|^2-|\tilde{F}_{1,n}|^2) \\
    &=-\kappa e^2 \int_0^\infty dq e^{-q^2 l_B^2/2}\sum_{n=2}^\infty (\frac{1}{2}-\frac{1}{2} \cos{\theta^{(n)}})(|\tilde{F}_{0,n}|^2-|\tilde{F}_{1,n}|^2)
    \\
    &=- \frac{1}{2} \kappa e^2 \int_0^\infty dq e^{-q^2 l_B^2/2}\sum_{n=2}^\infty(|\tilde{F}_{0,n}|^2-|\tilde{F}_{1,n}|^2)+  \kappa e^2 \int_0^\infty dq e^{-q^2 l_B^2/2}\sum_{n=2}^\infty \frac{1}{2} \cos{\theta^{(n)}}(|\tilde{F}_{0,n}|^2-|\tilde{F}_{1,n}|^2) 
    \\
    &= \Delta^X_0+\delta\Delta^X
\end{align}
where we recognize the first term as $\Delta^X_0$, and now focus on evaluating the second term which we denote as $\delta\Delta^X$. To do so, we write explicitly in terms of the Laguerre polynomials. For the $\Sigma_0$ term, we have:
\begin{align}
    \sum_{n=2}^\infty |\tilde{F}_{0,n}|^2 c_n = \sum_{n=0}^\infty c_n \frac{1}{n!}(\frac{k^2}{2})^n (L_0^n(k^2/2))^2
\end{align}
where we have abbreviated $\frac{1}{2}\cos{\theta^{(n)}} \equiv c_n $ and have extended the sum to include $n=0,1$. Since $c_n$ is not defined for $n<2$, we can set it to zero to recover the initial sum. Also note here $k$ is measured in units of $1/l_B$. At the end we will re-introduce the factors of $l_B$. 
%If we want to recover the particle-hole symmetric obtained above, we can set $c_0=c_1=1/2$, and do the same addition and subtraction. 
Since $L_0^n=1$,for the $n=0$ sum we simply have: 

\begin{equation}
     \sum_{n=0}^\infty c_n \frac{1}{n!}\left(\frac{k^2}{2}\right)^n %\xrightarrow[]{c_n = \frac{1}{2}} \frac{1}{2} e^{k^2/2}
\end{equation}

The $n=1$ sum is a bit more complicated. We first take out the $n=0$ term because the expression for $F_{1,n}$ is defined for $n\ge1$:
\begin{align}
    \begin{split}
        \sum_{n=0}^\infty |\tilde{F}_{1,n}|^2 c_n=\sum_{n=1}^\infty c_n \frac{1}{n!}(\frac{k^2}{2})^{n-1} (L_1^{n-1}(k^2/2))^2+c_0 k^2/2
    \end{split}
\label{sum1}
\end{align}
We then substitute the expression for the Laguerre polynominal $L_1^{n-1}(k^2/2)=1+(n-1)-k^2/2=n-k^2/2$. So \eqref{sum1} becomes
\begin{align}
    \sum_{n=1}^\infty c_n \frac{1}{n!}(\frac{k^2}{2})^{n-1} [n^2+(k^2/2)^2-k^2n] +c_0 k^2/2
\end{align}
We now look at each term in the square brackets separately. Starting with the second term we take one factor of $k^2/2$ outside, add and subtract 1 to get
\begin{align}
    \frac{k^2}{2}\left(\sum_{n-1}^\infty c_n \frac{1}{n!} (\frac{k^2}{2})^2+1-1 \right)= \frac{k^2}{2} \left( \sum_{n=0}^\infty c_n \frac{1}{n!} (\frac{k^2}{2})^n-1\right) %\xrightarrow[]{c_n=1/2} \frac{1}{2}\frac{k^2}{2}(e^{k^2/2}-1)
\label{Second}
\end{align}
For the last term in the square brackets, we cancel the factor of $n$ with the factorial to get
\begin{align}
    -\sum_{n=1}^\infty c_n k^2 \frac{1}{(n-1)!} (\frac{k^2}{2})^{n-1}=-\sum_{\tilde{n}=0}^\infty c_{\tilde{n}+1} k^2 \frac{1}{(\tilde{n})!} (\frac{k^2}{2})^{\tilde{n}} %\xrightarrow[]{c_n=1/2} -\frac{1}{2}k^2 e^{k^2/2}
\label{Third}
\end{align}

Now for the first term in the square brackets, we first cancel one factor of n with the factorial, then add and subtract 1, to get 
\begin{align}
    \sum_{n=1}^\infty c_n \frac{n^2}{n!} (\frac{k^2}{2})^{n-1}=\sum_{n=1}^\infty c_n \frac{(n-1)+1}{(n-1)!} (\frac{k^2}{2})^{n-1}
\end{align}
which we can split into 2 sums: 
\begin{align}
\begin{split}
    &\sum_{n=2}^\infty c_n \frac{1}{(n-2)!}\left(\frac{k^2}{2}\right)^{n-1}+\sum_{n=1}^\infty c_n \frac{1}{(n-1)!}  \left(\frac{k^2}{2}\right)^{n-1}\\
    &\sum_{\tilde{n}=0}^\infty c_{\tilde{n}+2} \frac{1}{\tilde{n}!} \left(\frac{k^2}{2}\right) \left(\frac{k^2}{2}\right)^{\tilde{n}}+\sum_{\tilde{n}=0}^\infty c_{\tilde{n}+1} \frac{1}{\tilde{n}!} \left(\frac{k^2}{2}\right)^{\tilde{n}} %\xrightarrow[]{c_n=1/2} \frac{k^2}{2} e^{k^2/2}+e^{k^2/2}
\end{split}
\label{First}
\end{align}

Putting everything together, we get:
\begin{align}
\tilde{\Sigma}_1-\tilde{\Sigma}_0&=\sum_{n=0}^{\infty} \frac{1}{n !}\left(\frac{k^2}{2}\right)^n\left[c_{n+1}-c_n\right]+\sum_{n=0}^{\infty} \frac{1}{n !} k^2 \cdot\left(\frac{k^2}{2}\right)^n\left[\frac{1}{2} c_{n+2}+\frac{1}{2} c_n-c_{n+1}\right] \\
&=\sum_{n=0}^{\infty} \frac{1}{n !}\left(\frac{k^2}{2}\right)^n\left[c_{n+1}-c_n\right]+\sum_{n=0}^{\infty} \frac{1}{n !}  \cdot\left(\frac{k^2}{2}\right)^{n+1}\left[ c_{n+2}+c_n- 2c_{n+1}\right]\\
&=\sum_{n=0}^{\infty} \frac{1}{n !}\left(\frac{k^2}{2}\right)^n\left[c_{n+1}-c_n\right]+\sum_{\tilde{n}=1}^{\infty} \frac{1}{(\tilde{n}-1) !}  \cdot\left(\frac{k^2}{2}\right)^{\tilde{n}}\left[ c_{\tilde{n}+1}+c_{\tilde{n}-1}- 2c_{\tilde{n}}\right] \\
&=\sum_{n=0}^{\infty} \frac{1}{n !}\left(\frac{k^2}{2}\right)^n\left[c_{n+1}-c_n\right]+\sum_{\tilde{n}=1}^{\infty} \frac{\tilde{n}}{(\tilde{n}) !}  \cdot\left(\frac{k^2}{2}\right)^{\tilde{n}}\left[ c_{\tilde{n}+1}+c_{\tilde{n}-1}- 2c_{\tilde{n}}\right] \\
&=c_1-c_0+\sum_{n=1}^{\infty} \frac{1}{n !}\left(\frac{k^2}{2}\right)^n\left[ (n+1) c_{n+1} + n c_{n-1} - (2n+1) c_n  \right]\\
&=\sum_{n=1}^{\infty} \frac{1}{n !}\left(\frac{k^2}{2}\right)^n\left[ (n+1) c_{n+1} + n c_{n-1} - (2n+1) c_n  \right]
\label{deltaSigma}
\end{align}
where $\tilde{\Sigma}_i(k)$ by $\Sigma_i \equiv \int_0^\infty dk e^{-k^2/2} \tilde{\Sigma}_i(k)$. What is left is evaluating the integrals using:

\begin{equation}
    \int_0^\infty dq e^{-q^2/2} (\frac{q^2}{2})^n= \frac{1}{\sqrt{2}} \Gamma(n+\frac{1}{2})
\end{equation}

and reintroducing the factor of $l_B$, we get:
\begin{equation}
    \delta \Delta^X=\frac{\kappa e^2}{ \sqrt{2} l_B}\sum_{n=1}^{\infty} \frac{1}{n !} \Gamma(n+\frac{1}{2}) \left[(2n+1) c_n -(n+1) c_{n+1} - n c_{n-1}  \right] 
    \label{deltaDeltaSM}
\end{equation}

In the limit of $n \rightarrow \infty$ the $\sigma_z$ term in the single particle Hamiltonian becomes small compared to the $\sigma_x$ term and the density matrix reduces to the particle-hole symmetric case for which $c_n \propto \cos{(\theta^{(n)})} \approx 0$ and the sum dies off giving a convergent result.
\subsection{Parametric Dependence}
Having the closed form expression for $\delta\Delta^X$ \eqref{deltaDeltaSM} allows us to examine its dependence on model parameters. All parameter dependence enters through:

\begin{align}
\begin{split}
    \cos{(\theta^{(n)}})&=\frac{\tilde{\gamma}^\prime/2}{\sqrt{(\tilde{\gamma}^\prime/2)^2+ \tilde{\gamma}_0^2 (n-1)n}}=
    \frac{\tilde{\gamma}^\prime/2}{|(\tilde{\gamma}^\prime/2)|\sqrt{1+ \frac{\tilde{\gamma}_0^2}{(\tilde{\gamma}^\prime/2)^2} (n-1)n}}\\&=
    \mathrm{sgn}(\tilde{\gamma}^\prime)\frac{1}{\sqrt{1+ \frac{\tilde{\gamma}_0^2}{(\tilde{\gamma}^\prime/2)^2} (n-1)n}} \approx \frac{1}{\sqrt{r}} \frac{1}{\sqrt{n(n-1)}}
\end{split}
\end{align}

where we defined the dimensionless ratio $r\equiv \frac{\tilde{\gamma}_0^2}{(\tilde{\gamma}^\prime/2)^2} \approx 184$  for the parameter values given in \cite{Jeil}, so 

\begin{align}
    \delta \Delta^X\sim \frac{1}{\sqrt{r}}=|\frac{(\tilde{\gamma}^\prime/2)}{\tilde{\gamma}_0}|= |\frac{t_4}{t_0}-\frac{\Delta^\prime}{2 t_1}| \sim \frac{\gamma^\prime}{\gamma_0}
\end{align}
\subsection{Orbital Coherence}
To calculate the energy of a state where a coherent superposition $\ket{\Psi}= \cos{(\theta/2})\ket{0} + e^{i \phi} \sin{(\theta/2)}\ket{1}$ is occupied for each state within a LL, we first calculate the the exchange Hamiltonian \eqref{HX} for the corresponding density matrix:
\begin{equation}
\rho=\left(\begin{array}{cc}
\cos ^2(\theta / 2) & \cos (\theta / 2) \sin (\theta / 2) e^{i \phi} \\
\cos (\theta / 2) \sin (\theta / 2) e^{-i \phi} & \sin ^2(\theta / 2)
\end{array}\right) \delta_{X,X'}.
\end{equation}

This gives 
\begin{equation}
H_X=-\frac{\kappa e^2}{l_B}\left(\begin{array}{cc}
X_{00} \rho_{00}+X_{01} \rho_{11} & \chi \rho_{01} \\
\chi \rho_{10} & X_{11} \rho_{11}+X_{10} \rho_{00}
\end{array}\right)
\end{equation}
where we denote the exchange integrals by
\begin{align}
\begin{split}
    &\int_0^\infty dq |F_{0,0}|^2 =X_{00}=\sqrt{\frac{\pi}{2}}, \\
    &\int_0^\infty dq |F_{1,1}|^2 =X_{11}=\frac{3}{4}\sqrt{\frac{\pi}{2}},\\
    &\int_0^\infty dq |F_{0,1}|^2  =X_{01}=X_{10}=\frac{1}{2}\sqrt{\frac{\pi}{2}},\\
    &\int_0^\infty dq F_{0,0} F_{1,1}^*=\chi =\frac{1}{2}\sqrt{\frac{\pi}{2}}.
\end{split}
\label{exchange_integrals}
\end{align}
We can then calculate the exchange energy
\begin{equation}
\frac{E_X}{N_\phi}=\frac{1}{2 N_\phi} \operatorname{Tr}\left\{H^X \cdot \rho\right\}=-\frac{1}{64} \frac{\kappa e^2}{l_B} \sqrt{\frac{\pi}{2}}(-4 \cos (\theta)+\cos (2 \theta))
\end{equation}
where $N_\phi$ is the LL degeneracy that comes from summing over the intra-LL label. 

\begin{figure}[H]
\begin{center}
\includegraphics[width=\linewidth]{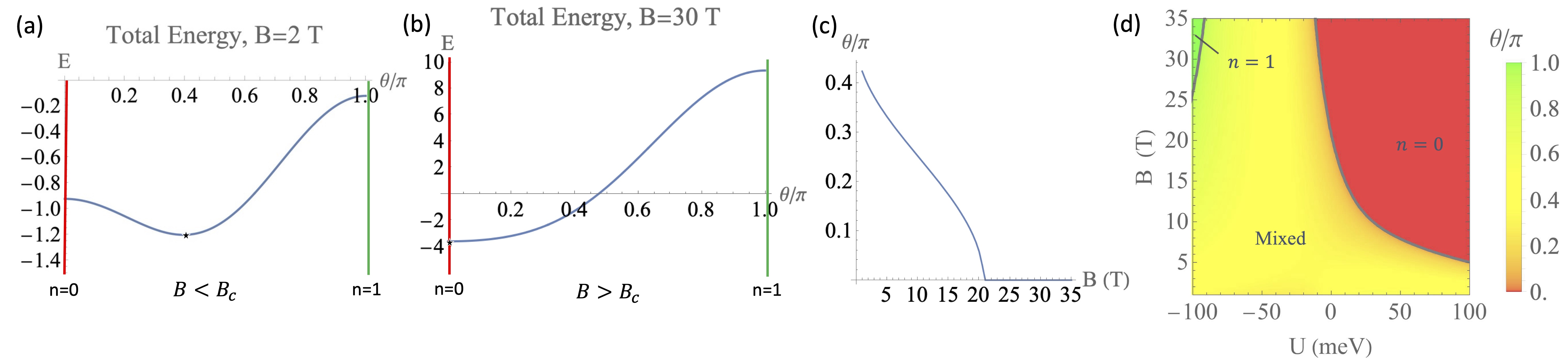}
\end{center}
\caption{(a-b) Total energy as a function of the polar angle $\theta$ such that 0 indicates occupying the $n=0$ LL, and $\theta/\pi=1$ indicates occupying the $n=1$ LL (a) at magnetic field where the one-body terms compete with the exchange self-energy , and (b) at magnetic fields where both the one-body terms and the exchange self-energy favor $n=0$. (c) Polar angle $\theta_{\mathrm{min}}$ at which the total energy is minimized as $B$ is varied at $U=0$. 
(d) Phase diagram for the spinless valley-less limit at half-filling. Color scale represents the polar angle $\theta_{\mathrm{min}}/\pi$, $\epsilon_r=5$. %(c)  is Line cut of (a) showing $\theta_{\mathrm{min}}$ at $U=0$. 
}
\label{flavorless}
\end{figure}

The one body terms in the Hamiltonian, which receive contributions from both single particle physics and the Lamb shift:
\begin{equation}
T=\left(\begin{array}{cc}
0 & 0 \\
0 & \Delta_{S P}-\Delta^X
\end{array}\right)
\end{equation}
contribute to the total energy by an amount $\left(\Delta_{S P}-\Delta^X\right) \sin ^2(\theta / 2)$ per LL. %$\Delta \equiv \Delta_{SP}-\Delta^X$. 
In the particle-hole symmetric limit, when $\Delta_{SP}=0$ and $\Delta^X=\Delta^X_0=\frac{1}{8}\sqrt{\frac{\pi}{2}}$, the total energy $E_T \propto \cos{(2\theta)}$ and is therefore always minimized at $\theta=\pi/2$; that is, an equal superposition of $n=0$ and $n=1$. Away from this limit, the total energy can be minimized anywhere between $0 \le \theta \le \pi$. 
The results of mean field ground state are summarized in Fig. \ref{flavorless}.
 %Indeed, one can prove that should be the case in the presence of particle-hole symmetry [Insert argument?]

\subsection{Remote Level Gaps for Half-filled $n=1$ Landau Levels}
\begin{figure}[H]
\begin{center}
\includegraphics[width=0.7\linewidth]{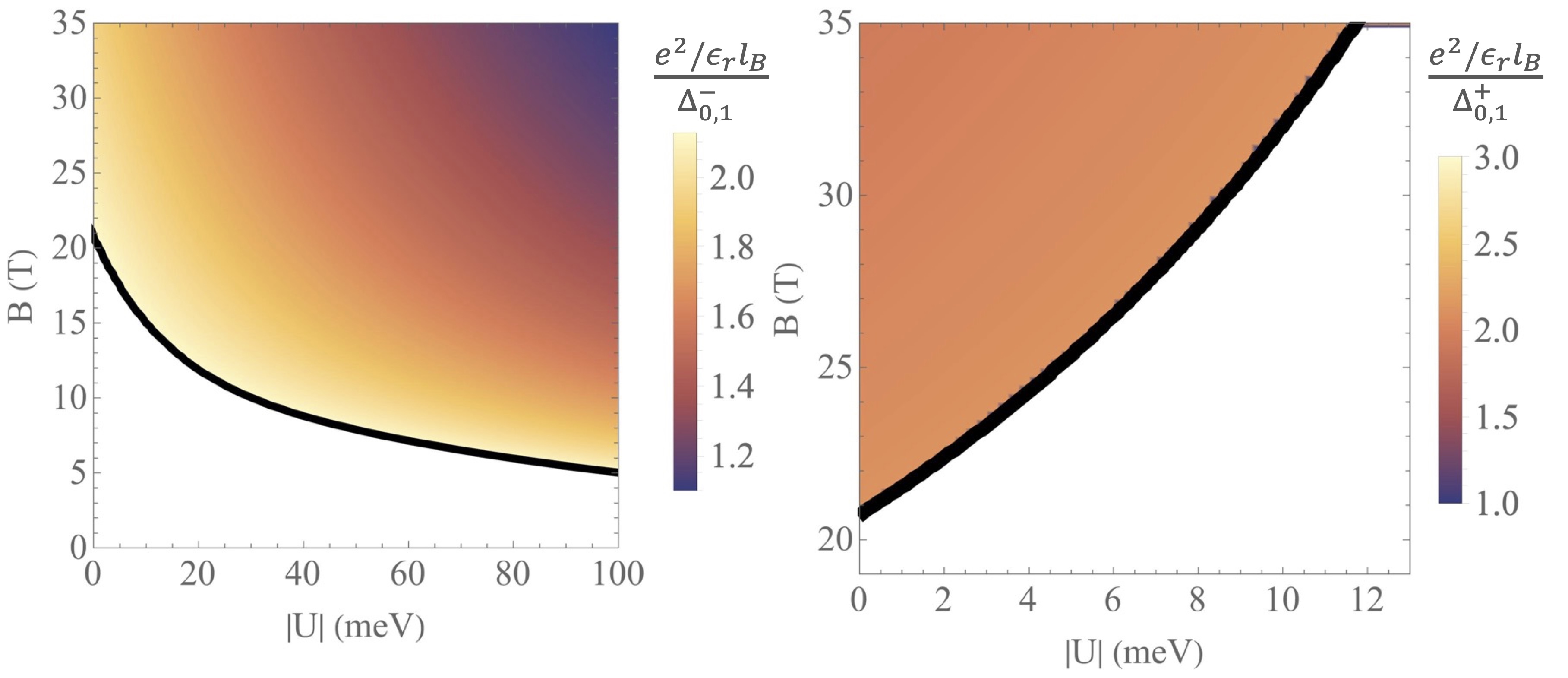}
\end{center}
\caption{Ratio of Coulomb interaction energy scale $e^2/\epsilon_r l_B$ to the gap between $n=0$ and $n=1$ with the $n=0$ LL fully occupied and the $n=1$ LL half-filled (Fig. \ref{Sector_Gap} in main text).
}
\label{mixing_par}
\end{figure}

To obtain the energies of adding an electron to a half-filled $n=1$ and remove it from a full $n=0$ orbital, we begin by calculating the mean field Hamiltonian within each spin/valley flavor sector:

\begin{equation}
H^{\tau,\sigma}=\left(\begin{array}{cc}
\epsilon_{\tau, \sigma} & 0 \\
0 & \epsilon_{\tau, \sigma}
\end{array}\right)+\left(\begin{array}{cc}
0 & 0 \\
0 & \Delta^{(\tau)}
\end{array}\right)-\frac{\kappa e^2}{l_B}\left(\begin{array}{cc}
v_0^{\tau,\sigma} X_{00}+v_1^{\tau,\sigma} X_{10} & 0 \\
0 & v_1^{\tau,\sigma} X_{11}+v_0^{\tau,\sigma} X_{10}
\end{array}\right)
\end{equation}

in the parameter regime obtained from Fig. \ref{flavorless} d, where the ground state is expected not to be orbitally mixed, so the mean field state will be given by a full $n=0$ LL and a half-filled $n=1$ LL ($\nu_0^{\tau,\sigma}=1, \nu_1^{\tau,\sigma}=1/2$ ), giving energy difference between the two orbitals:
\begin{align}
    \Delta_{01}^{(\tau)}&=\Delta^{(\tau)}- \frac{\kappa e^2}{l_B} (X_{11}/2+X_{10}/2 - X_{00}- X_{10}/2)\\
    &=\Delta^{(\tau)}- \frac{\kappa e^2}{l_B} (X_{11}/2+X_{10}/2-X_{00}),
\label{Sup_sector_Gap}
\end{align}
where $\Delta^{(\tau)}$ is the one-body splitting between $n=0$ and $n=1$ including the contribution due to the displacement potential for each valley $\tau$, $\epsilon_{\tau, \sigma}$ is the single-particle  spin and valley Zeeman shifts, and $X_{ij}$ are the exchange integrals defined in Eq. \eqref{exchange_integrals} . One way to assess the robustness against quantum fluctations is to compared $\Delta_{01}$ to the Coulomb interaction energy scale $e^2/\epsilon_r l_B$, as illustrated in Fig. \ref{mixing_par}. A larger ratio is expected reduce the gap of non-Abelian states \cite{PRX_2015_Mixing,luo2016missing, KunYang_Review}.

\begin{figure}
%\begin{center}
\includegraphics[width= \linewidth]{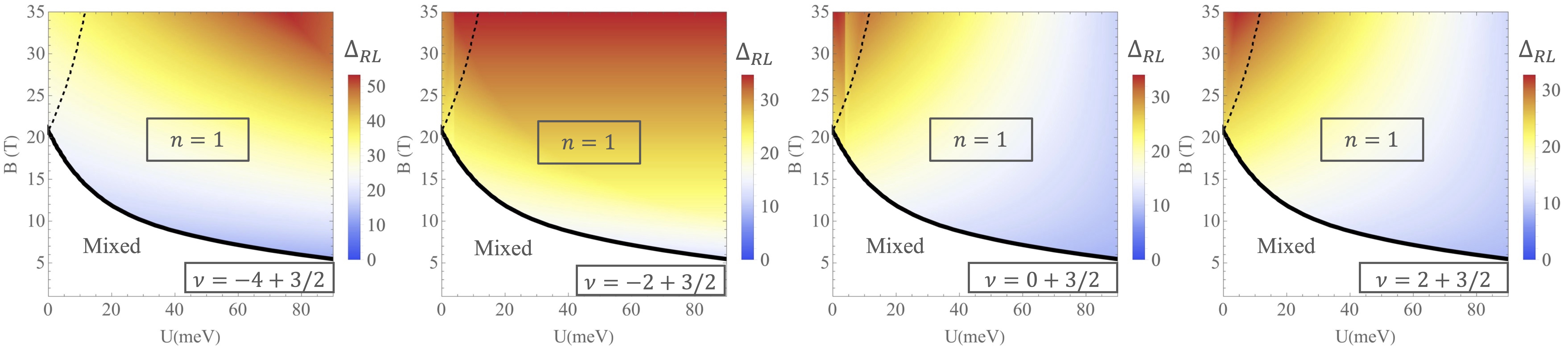}
%\end{center}
\caption{Gap to remote levels when the $n=1$ LL is half-filled, for $\epsilon_r=5$, and constant Zeeman field $B_\mathrm{ZM}=35$ T.  }
\label{QPEs}
\end{figure}

Additionally, the gap to remote levels regardless of spin and valley is evaluated at different filling factors by calculating the minimum energy to add or remove an electron from the half-filled valence $n=1$ LL, which is illustrated in Fig. \ref{QPEs}. For $\nu=-2+3/2$, there is a transition at small $U$ when it becomes larger than the spin Zeeman splitting. Before the transition the both valleys are occupied for the majority spin, while after the transition both spins are occupied for the majority valley. Right after the transition (at intermediate $U$) the closest level is $n=0$ of the same flavor, while for larger $U$, the closest level becomes the filled $n=1$ of the majority-valley majority-spin, thus limiting the gap to a $U$ independent value at large $U$. For positive filling factors, the valence $n=1$ level is in the high energy valley (except at $\nu=3/2$, when $U$ is smaller than the spin Zeeman splitting), so its energy is lowered by the displacement field, leading to the mixed state at odd filling factors beyond the dashed boundary in Fig. \ref{QPEs}.

\end{document}